\documentstyle[aas2pp4]{article}

\begin{document}

\title{A Search for Environmental Effects on Type Ia Supernovae}
\author{Mario Hamuy}
\affil{Steward Observatory, The University of Arizona, Tucson, AZ 85721} 
\authoremail{mhamuy@as.arizona.edu}
\author{S. C. Trager\altaffilmark{1}}
\altaffiltext{1}{Hubble Fellow}
\affil{The Observatories of the Carnegie Institution of Washington,
813 Santa Barbara St., Pasadena, CA 91106}
\authoremail{sctrager@ociw.edu}
\author{Philip A. Pinto}
\affil{Steward Observatory, The University of Arizona, Tucson, AZ 85721}
\authoremail{ppinto@as.arizona.edu}
\author{M. M. Phillips}
\affil{Carnegie Institution of Washington, Las Campanas Observatory,
Casilla 601, La Serena, Chile} 
\authoremail{mmp@lco.cl}
\author{R. A. Schommer}
\affil{National Optical Astronomy Observatories\altaffilmark{2}, Cerro
Tololo Inter-American Observatory, Casilla 603, La Serena, Chile}
\altaffiltext{2}{Cerro Tololo Inter-American Observatory, Kitt Peak
National Observatory, National Optical Astronomy Observatories,
operated by the Association of Universities for Research in Astronomy,
Inc., (AURA), under cooperative agreement with the National Science
Foundation.}
\authoremail{rschommer@noao.edu}
\author{Valentin Ivanov}
\affil{Steward Observatory, The University of Arizona, Tucson, AZ 85721}
\authoremail{vivanov@as.arizona.edu}
\author{Nicholas B. Suntzeff}
\affil{National Optical Astronomy Observatories\altaffilmark{2}, Cerro
Tololo Inter-American Observatory, Casilla 603, La Serena, Chile}
\authoremail{nsuntzeff@noao.edu}

\begin{abstract}
We use integrated colors and $B$ and $V$ absolute magnitudes of Type
Ia supernova (SN) host galaxies in order to search for environmental
effects on the SN optical properties. With the new sample of
44 SNe we confirm the conclusion by Hamuy et al. (1996a) that bright
events occur preferentially in young stellar environments. We find
also that the brightest SNe occur in the least luminous  galaxies, a possible
indication that metal-poorer neighbourhoods produce the more luminous
events. The interpretation of these results is made difficult,
however, due to the fact that galaxies with younger stellar
populations are also lower in luminosity. In an attempt to remove this ambiguity we
use models for the line strengths in the absorption spectrum of five
early-type galaxies, in order to estimate metallicities and ages of the
SN host galaxies.  With the addition of abundance
estimates from nebular analysis of the emission spectra of three spiral
galaxies, we find possible further evidence that luminous SNe are produced in
metal-poor neighborhoods.  Further spectroscopic observations of the
SN host galaxies will be necessary to test these results and
assist in disentangling the age/metallicity effects on Type Ia SNe.

\end{abstract}


\section{Introduction}

The study of the Hubble diagram of nearby ($z\leq0.1$) Type Ia
supernovae (SNe, hereafter) has shown that these objects can be used
as precise distance indicators through the application of the peak
luminosity-light curve width relation
(\cite{phillips93,hamuy96b,riess96,perlmutter99}).  This method has
been recently applied to distant ($0.3\leq z\leq1$) SNe to measure the
values of the cosmological parameters. These studies have revealed a
surprising cosmological result, namely, that the expansion of the
Universe is presently accelerating due to a non-zero cosmological
constant (\cite{riess98,perlmutter99}).  Even though this conclusion
is based on several dozen of high-$z$ events that permit the lowering of
the statistical errors to significant levels, the reality of this
result is threatened by the systematic errors that may result from
comparison of distant events with those of the local
Universe. Particularly relevant is the evolution of the stellar
populations that produce the high-$z$ SNe (with look-back times of 5-7
Gyr) which has the potential to affect the properties of the explosion
and bias the determination of distances.  It proves important
therefore to investigate these effects from theoretical and
observational point of views.

The effects of ages and metallicities of the SN progenitors in the
properties of the SNe have been examined from a theoretical point of
view using delayed detonation (DD) models of a Chandrasekhar-mass
white dwarf (WD) by H\"oflich et al. (1998), who concluded that 1) old
stellar populations contain WDs with C/O=1/1 that produce SNe with
somewhat fainter peak luminosities ($\sim0.1$ mag) relative to SNe in
younger environments containing WDs with C/O=2/3, and 2) a drop of 0.5
dex in the metallicity of the SN progenitor leaves the light curve
shapes virtually unchanged and causes a slight decrease (0.03 mag) in the
monochromatic $B$ and $V$ luminosities.  Using DD models and the ``strong
wind'' model for the mass-accretion phase of the WD, Umeda et
al. (1999) reached the same conclusion, namely, that lower metallicity
or older progenitors lead to dimmer SNe Ia, although the metallicity
effects are more pronounced in these models due to opacity effects in
the mass-accretion rate.

Since the local Universe produces SNe that explode in different
stellar environments, the sample of nearby SNe offers
the possibility to test these
theoretical predictions and examine the origin of the diversity among
SNe Ia.  A first attempt using this approach based on the
morphological classification of the host galaxies was performed by
Hamuy et al. (1995, 1996a), who noticed that the most luminous SNe Ia were
hosted by spiral (blue) galaxies. Using the Hamuy et al. (1995) sample
of 13 distant SNe Ia and 10 nearby SNe, 
Branch et al. (1996) reached a similar conclusion, namely, that SNe in
redder galaxies have lower luminosities. The goal of this paper is to go a step
further and use all the information available to us on the SN host
galaxies, in order to collect observational clues that could help us
unveil the nature of SNe Ia and the relevant parameters that govern
their optical properties.  Ideally, a study of the effects of the
stellar environments on the SN Ia explosions would require
observations of the immediate stellar population that produced the SN
but, in practice, this is difficult to perform.  A more feasible
observation, on the other hand, is that of the integrated light of the
galaxy through an aperture containing a significant fraction of the
total luminosity of the galaxy. Although this observation proves much
easier to perform, the global properties of the stellar system might
differ significantly from those of the site where the SN progenitor
formed.  An additional difficulty of this approach that hampers the
interpretation of the results comes about from the observations of
present-day stellar populations which might be significantly different
than those where the SN progenitor was born. Given that the direct
observation of a SN progenitor is unlikely, we proceed
here with this indirect route, keeping in mind its limitations
and the potential danger of extrapolating the nature of the host galaxy
globally to the specific nature of the SN environment.

In this paper we focus both on photometric and spectroscopic data of
the 62 galaxies that have hosted the best observed SNe.  Section 2
describes the object sample and the photometric data that we employ to
characterize the SN neighborhoods. In Section 3 we use this
information and additional limited spectroscopic data available in the
literature to study the relation of the SNe to the ages and
metallicities of the host galaxies. In Section 4 we compare the
observational results to the theoretical predictions, and we analyze
the systematic effects of the environmental conditions in the
determination of extragalactic distances. Finally, in Section 5 we
summarize our conclusions.

\section{Supernova Sample and Observations}

In this paper we consider the 62 best-observed SNe Ia with $z<0.1$
comprising the full sample of 29 Cal\'an/Tololo SNe (\cite{hamuy96c}),
20 objects from the ``CfA'' sample (\cite{riess99}), and 13 nearby SNe
Ia (SNe 1937C, 1972E, 1980N, 1981B, 1986G, 1989B, 1990N, 1991T, 1992A,
1994D, 1996X, and 1996bu). This is the same sample studied by Phillips
et al. (1999) having precise $BVI$ peak magnitudes, decline rates
[$\Delta m_{15}(B)$], and color excesses, due to absorption in the host
galaxy.  Table 1 gives the full list of SNe, their decline rates, the
name and Hubble type of the host galaxy. The morphological
classifications come from 1) our own typing of the Cal\'an/Tololo SN
host galaxies\footnote[1] {In three cases we use Hubble Space Telescope images secured
through the ``Supernova Host Galaxies'' Cycle 6 SNAP program GO-6362.
For the rest of the sample we employ ground-based CCD frames.}, 2) the NASA/IPAC Extragalactic
Database\footnote[2] {The NASA/IPAC Extragalactic Database (NED) is
operated by the Jet Propulsion Laboratory, California Institute of
Technology, under contract with the National Aeronautics and Space
Administration.}  (http://nedwww.ipac.caltech.edu/), and 3) Riess et
al. (1999), in this order of preference.

One of the easiest measurables of a galaxy is its total luminosity.
For 25 of the Cal\'an/Tololo SN host galaxies we use our own CCD
images (obtained with the CTIO Blanco 4-m or the 0.9-m telescope) to
calculate instrumental magnitudes through concentric circular
apertures (after removing foreground stars from the aperture). We
perform the transformation of these magnitudes to the standard $BV$
system using the photometric sequences around these galaxies
(\cite{hamuy96c}), and the corresponding color terms for the CTIO CCD
cameras. For each galaxy we estimate its total magnitude from the
curve of growth obtained. Since most of these galaxies lie well inside
the Hubble flow we apply then K corrections that we compute 
from our own spectra of these galaxies. The next step is to
correct these magnitudes for foreground reddening, which we perform
using the values from Schlegel et al. (1998).  Since there is no
universally accepted method to correct the galaxy luminosities for
internal absorption, we skip this correction.  For 19 galaxies of the
sample integrated magnitudes are readily available from the Third
Reference Catalogue of Bright Galaxies (\cite{devaucou91}, hereafter
RC3).  For consistency with our measurements we choose from RC3 the
integrated magnitudes $B_T$ and $V_T$, which are uncorrected for
internal absorption.  Table 1 (columns 5-6) lists the integrated $B_T$
and $V_T$ magnitudes for such galaxies and the Cal\'an/Tololo SN
hosts. Also given in columns 7 and 8 are the absolute $B$ and $V$
magnitudes of the host galaxies, which we compute as follows. For the
hosts of the nearby ($cz\leq1000$ km sec$^{-1}$) SNe 1937C, 1972E,
1981B, 1989B, 1990N, and 1998bu we employ the Cepheid distances
measured by Gibson et al. (2000). For the early-type hosts of the
nearby SNe 1980N, 1986G, 1991bg, 1992A, and 1994D we use the SBF/PNLF
distances listed in Hamuy et al. (1996a). For all of the other hosts
with $cz\geq2000$ km sec$^{-1}$ we use the recession velocity to compute the
distance modulus, except for the host of SN 1996ai, whose redshift of 951 km sec$^{-1}$
is low enough to be affected by peculiar motion.
To match the Cepheid distance scale of Gibson et
al. (2000) we adopt $H_0$=69 km sec$^{-1}$ Mpc$^{-1}$.

\section{Results}

In this section we use the observations of the SN Ia host galaxies to
explore the effect of the stellar environment on the SNe.
To characterize a SN we use the decline rate, $\Delta m_{15}(B)$, which
is defined as the amount that the SN declines in $B$ band magnitude
during the first 15 days after maximum light. It is a measure of
the post-maximum decline rate, with fast-declining SNe having
large $\Delta m_{15}(B)$ values and narrower light curves.
This parameter proves to be a convenient reddening-free and distance-free 
estimate of the SN peak brightness (\cite{phillips99}), with
the peak luminosity decreasing with increasing decline rate.

\begin{figure}
\epsscale{1.0}
\plotone{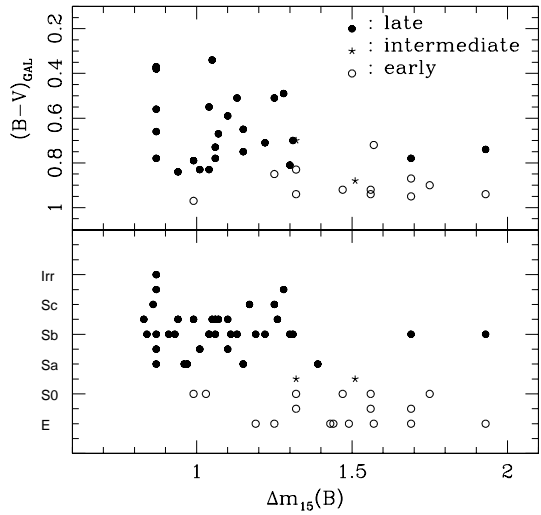}
\caption{(top) The $(B-V)$ color of the SN host galaxy
vs. the decline rate of the SN.  (bottom) The morphological type of
the SN host galaxy vs. the decline rate of the SN. \label{fig1}}
\end{figure}

The bottom panel of Figure \ref{fig1} shows the morphological-type
vs. $\Delta m_{15}(B)$ diagram for the objects of our sample. This
plot confirms the result found by Hamuy et al. (1996a), namely, that
elliptical galaxies have hosted only fast-declining SNe and that
spirals produce preferentially slow-declining SNe.  The top panel of
Figure \ref{fig1} shows the $B-V$ color of the host galaxies
vs. $\Delta m_{15}(B)$.  The remarkable feature of this ``L-shaped''
plot is the lack of fast-declining SNe among blue galaxies [$(B-V)\le0.7$].
Galaxies with $(B-V)\ge0.7$, on the other hand, have produced SNe 
with a wide range of decline rates, which is somewhat different than
the result of Branch et al. (1996) that revealed
a lack of slow-declining SNe among red hosts.
In any case, both diagrams point to the same conclusion, namely, that the distribution
of decline rates in elliptical galaxies is markedly different than
that in spirals, in the sense that younger environments produce
preferentially the slowest-declining (luminous) SNe (and vice versa).

\begin{figure}
\epsscale{1.0}
\plotone{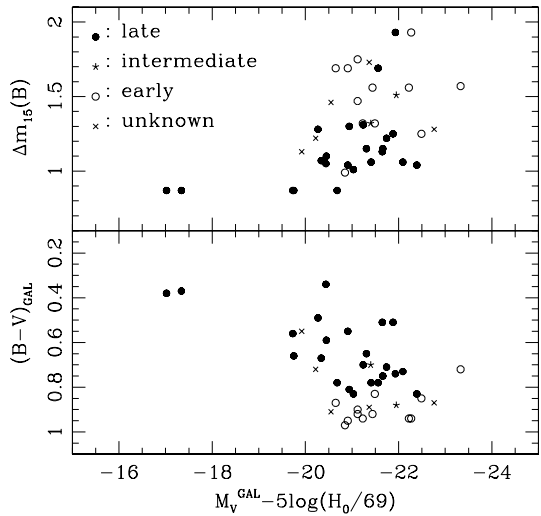}
\caption{(top) The decline rate of the SN vs. the
absolute $V$ magnitude of the SN host galaxy.  (bottom) The $(B-V)$
color vs. the absolute $V$ magnitude of the SN host galaxy. \label{fig2}}
\end{figure}

The top panel of Figure \ref{fig2} shows the absolute $V$ magnitude of
the host galaxies vs. $\Delta m_{15}(B)$.  This plot reveals a trend
for late-type galaxies, in the sense that the slowest-declining
(brightest) events have been hosted by the least luminous galaxies.  It is
well known that the integrated luminosities of spirals are correlated
(with a large scatter) with the oxygen abundance (\cite{henry99}),
presumably due to the fact that more massive galaxies have retained
more of the metals ejected by SNe. Although the scatter in Figure
\ref{fig2} is substantial, this plot suggests that metal-poor
environments (least luminous galaxies) harbor the brightest SNe (``dim
galaxy-bright SN'').  However, as we show in the bottom panel of
Figure \ref{fig2}, it is the case that the least luminous galaxies of our
sample are also the bluest. It is possible then that the metallicity
effect suggested by the galaxy luminosities is instead an age
effect, as suggested by Figure \ref{fig1}.  Conversely, it is
plausible that the age effect inferred by the galaxy colors is just a
reflection of a metallicity effect.

\begin{figure}
\epsscale{1.0}
\plotone{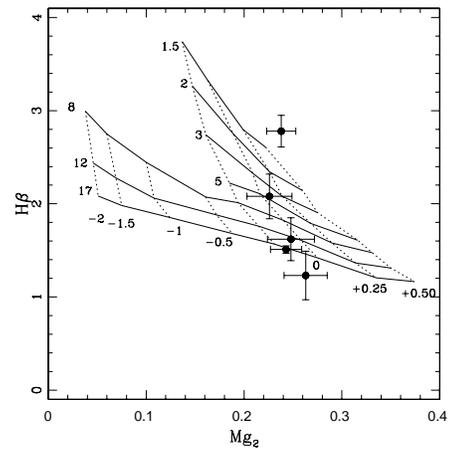}
\caption{H$\beta$ versus Mg$_2$ relations as a function
of age (1.5-17 Gyr) and metallicity ($-2\leq$[Fe/H]$\leq+0.5$)
from the population synthesis models of
Worthey (1994).  The galaxies listed in Table 2 are shown with solid
dots. \label{fig3}}
\end{figure}

What is clear from the integrated colors and luminosities of the SN Ia
host galaxies is that both age and/or metallicity affect the
luminosities of the SNe. It is important to try to disentangle these
effects and a promising route to break this degeneracy is to use
spectroscopic data for the host galaxies and the
stellar population synthesis models of Worthey (1994) for absorption
line strengths, at least in single-burst systems.  In the Worthey
models a single line index depends both on age and abundance but the
Mg$_2$-H$\beta$ plane (or other combinations of indices) permits, in
principle, the removal of the ``age-metallicity'' degeneracy.  Figure
\ref{fig3} shows such models for a wide range of ages (1.5-17 Gyr) and
abundances between [Fe/H]$=-2$ and +0.5.

To populate this diagram with SN host galaxies we used the
line strengths for the five early-type galaxies (NGC 1380, NGC 2962, NGC
4374, NGC 4526, and NGC 5061) observed by Trager et al.
(1998, hereafter T98)  at Lick Observatory with the IDS spectrograph
(with a small entrance aperture of $1\farcs4\times4\arcsec$).
High-quality line strengths for the central $r=r_e/8$ aperture of NGC
4374 were also obtained by Gonz\'alez (1993, hereafter G93) with the
IDS (and reproduced by \cite{trager00}; hereafter T00).  In the rest
of this paper we employ the G93 values for NGC 4374 and the T98
indices for the other cases. Since early-type galaxies display line-strength
gradients, it proves necessary to calculate aperture
corrections to convert the indices of T98, obtained through the
rectangular aperture of $1\farcs4\times4\arcsec$, to the equivalent of
the circular $r=r_e/8$ aperture of G93.  For this purpose we compute
synthetic line strengths through the circular and the rectangular
apertures, and the corresponding difference between them, which we add
to the observed index. The synthetic index is simply the
luminosity-weighted line strength inside the specified aperture, using
a $r^{1/4}$ law and the index gradients and zero-points from Davies et
al. (1993) for H$\beta$ and Mg$_2$, and from Kobayashi \& Arimoto
(1999) for Mgb, Fe5270, and Fe5335 (in actuality, since an aperture
correction is the difference between two synthetic line strengths, the
zero point of the index does not affect our results).  In all cases
these corrections prove to be $\sim15$\% or less, in the sense of
increasing H$\beta$ and decreasing the other indices.  Table 2 lists
the resulting absorption strengths for the five galaxies, which are
overplotted on the model grid of Figure \ref{fig3}.

The first thing to notice in this figure is that two  points lie
outside the parameter space. The departure of NGC 5061 from the model grid
can be simply understood as due to a young galaxy with high metal abundance.
NGC 2962 poses more problems since the weak H$\beta$ value
implies an age significantly larger than the Hubble time. It is possible
that the weak H$\beta$ value is due to contamination by H$\beta$
emission. Emission is present near the center of $\sim50$\% of
early-type galaxies (\cite{davies93}) and may substantially bias the
H$\beta$ absorption strength, especially on observations made through
small apertures.  A statistical correction based on [\ion{O}{3}]
emission has been proposed (T00), but these values are not available
for these galaxies. This problem can be overcome in the future with
further observations of these galaxies through bigger apertures, to
minimize the contamination by emission.  With this caveat
about the available data, we proceed with the determination of ages and
abundances for these galaxies.

In principle we can use the Worthey line strength models to determine
ages and metallicities. However, it is a well-known fact that the
abundances derived from the Mg and the Fe indices yield [Mg/Fe] ratios
which exceed the solar values by 0.2--0.3 dex (\cite{worthey92}). To
account for this effect in what follows we employ an extension of the
Worthey models that incorporates non-solar abundances (T00), in order
to derive single-stellar-population (SPP) equivalent ages,
metallicities, and enhancement ratios, [E/Fe], from the H$\beta$, Mgb,
and $\langle$Fe$\rangle$ line strengths.  Table 3 summarizes these
values for the five early-type galaxies of our sample.  The SPP ages span
a wide range between 24 and 1.4 Gyr, in agreement with the finding of
T00 from the sample of elliptical galaxies observed by G93.  We find
that the SPP abundances for the five galaxies span the range from
[Z/H]$=-$0.23 to +0.56 which, again, is in fair agreement with the
conclusion by T00 that early-type galaxies cover the range
$-0.1\leq$[Z/H]$\leq+0.7$. The enhancement ratios we derive here are 
also matched well by the strongly peaked ratios around
 $\langle$[E/Fe]$\rangle=+0.20$ found by T00 from the G93 sample.

\begin{figure}
\epsscale{1.0}
\plotone{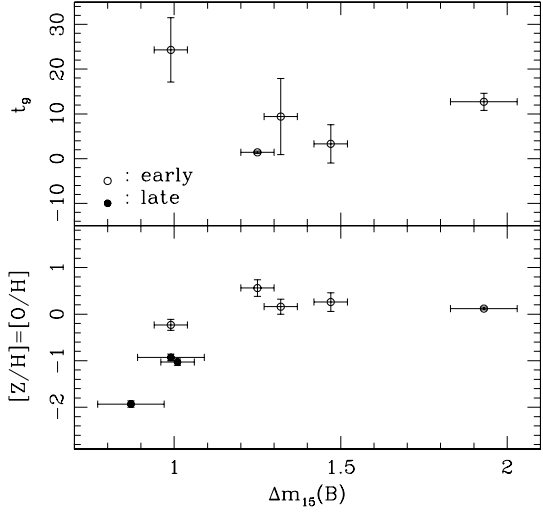}
\caption{(top) The age (in units of Gyr) of the stellar
populations of the SN host galaxies versus the decline rate of the
SN. (bottom) Metallicity of the stellar populations of the SN host
galaxies versus the decline rate of the SN. \label{fig4}}
\end{figure}

The top panel of Figure \ref{fig4} shows $t_9$ vs. $\Delta m_{15}(B)$,
where $t_9$ is the SPP age in Gyr.
Clearly the large error bars (mainly caused by the uncertainties in
H$\beta$) and the small number of objects in this plot prevent us from
drawing strong conclusions. Evidently, to be able to confirm or rule
out an age effect on SNe Ia from this diagram it will be necessary in
the future to secure high signal-to-noise spectra of as many
early-type SN host galaxies as possible.  The bottom panel of Figure
\ref{fig4} shows with open circles the [Z/H] values for the five SN Ia
host galaxies vs. $\Delta m_{15}(B)$. Note that in this case the
uncertainties in [Z/H] (mainly determined by the strong Mg feature)
are quite low (typically 0.1--0.2 dex) which permit us to say that there
appears to be a mild trend of metallicity with $\Delta m_{15}(B)$. Again,
the low number of objects renders it difficult to draw any strong
conclusion. 

To populate this plot with more points we used the SIMBAD Astronomical
Database to search for bibliographical references about metallicity
measurements of the SN Ia spiral hosts listed in Table 1.  This
search yielded abundances for only three galaxies, namely, NGC 3368, NGC
5005, and NGC 5253. For NGC 5253 we adopt the value of Kobulnicky et
al. (1999), log(O/H)$=-3.86\pm0.02$, from nebular analysis of the
integrated spectrum of the galaxy because this value represents a
global measurement. Note that this value is in close agreement with
the measurements of Storchi-Bergmann et al. (1994) [log(O/H)$=-3.6$]
and Campbell et al. (1986) [from log(O/H)$=-3.64$ to $-$3.83].  For
NGC 3368 we adopt the value at the effective radius $r_e$ (11.2 kpc)
quoted by Oey \& Kennicutt (1993), log(O/H)$=-2.96\pm0.02$, from
nebular analysis of three HII regions; this value is in good agreement
with the analysis of 25 HII regions by Dutil \& Roy (1999) who found a
central abundance of log(O/H)$=-3.02$ and a shallow abundance gradient
of $-0.009$ dex kpc$^{-1}$. Finally, for NGC 5005 we adopt the
characteristic abundance log(O/H)$=-2.86\pm0.02$ at $r_e$ (7.6 kpc)
determined by Oey \& Kennicutt (1993).  Adopting the solar abundance
(by mass) ratio log(O/H)$_{\odot}=-1.93\pm0.07$ (\cite{grevesse96}) we
obtain the [O/H] abundances for the three spirals of our sample which we
include in Table 3.

To allow a comparison between the [O/H] values of the late-type
galaxies and the [Z/H] abundances of the early-types we simply assume
[Z/H]=[O/H] for the latter, since O dominates Z in the line strength
models.  With the inclusion of the three spirals (shown as filled dots in
Figure \ref{fig4}) there appears to be somewhat more evidence for a
metallicity-decline rate relation for SNe Ia.  There are a few caveats
to keep in mind when examining this diagram. First, there are only 8
points in this diagram; second, the metallicities derived for spirals
from HII regions studies are snapshots of the present-day abundance of
the ISM in these galaxies, and it is likely that the metallicities of
the SNe progenitors were even lower (since they formed earlier);
third, the technique of nebular analysis in late-type galaxies and the
method of absorption line indices in early-type galaxies for abundance
determinations have not been tested on the same objects; and, fourth,
we use global abundances for the host galaxies and not the values at
the SN positions.

\section{Discussion}

To recapitulate, the photometric data of the SN Ia host galaxies show
clear evidence that there are environmental conditions that affect the
optical properties of SNe Ia. It is difficult though to identify
whether the age or the metallicity of the stellar populations is
responsible for the SN diversity, because the bluest galaxies of our
sample are also the least luminous.  In an attempt to disentangle the effects
of age and metallicity we estimate SPP parameters for the five early-type
galaxies with line strengths available in the literature, and we
collect metallicities from nebular analysis of the emission spectra of
three late-type galaxies. The spectroscopic data show that the age of the
stellar population is not correlated with $\Delta m_{15}(B)$ and that
metal-poorer environments produce the most luminous SNe. Given the low
number of galaxies in these diagrams and the large uncertainties in
the SPP ages we must take these results with caution. In any case, the
spectroscopic studies of the SN host galaxies provide a promising
route to unveil the cause of the SNe Ia diversity, and it is important
to start securing spectrophotometry of all SN host galaxies, both for
a nebular analysis of the integrated emission spectrum of late-type
galaxies and for a study of the absorption lines of the integrated
spectrum of early-type galaxies.

If age was the main parameter responsible for the range of optical properties
of SNe Ia, the photometric data suggest that younger
environments produce preferentially slow-declining (luminous) SNe, which 
agrees with the theoretical predictions of H\"oflich et al. (1998).
We will need to wait for further spectroscopic data, however, before we
can confirm or rule out the role of age as the main SN parameter.

Despite the difficulties in identifying the driving parameter in the SN
optical properties, the observations show that SNe Ia can occur in
environments with a wide range of metallicities. This observation
poses serious problems to the ``strong wind'' models advocated by
Umeda et al. (1999) which predict the absence of SNe Ia in metal-poor
environments.  According to these models galaxies with
[Fe/H]$\leq-1.0$ do not have WDs capable of blowing sufficiently
strong winds to allow the WD mass to grow to the Chandrasekhar
mass. This is contradicted by SN 1972E that went off in a very metal-poor
environment ([Z/H]$\sim-2$). These models are also challenged by
SNe 1996bi and 1998bu that exploded in galaxies with [Fe/H]$\sim-1.0$.

Given the present state of affairs and the difficulty to quantify the
weights of these parameters in the SN properties, it is interesting to
mention other pieces of observational data that could help us
understand the nature of SNe Ia. Since galaxies, in general, display
age and metallicity gradients, the study of the radial distribution of
SNe in the host galaxies proves to be a useful tool for this purpose.
According to the study by Wang et al. (1997), SNe Ia at smaller
galactocentric distances have larger  fluctuations in brightness than
those farther out.  Riess et al. (1999) reached the same conclusion
using peak magnitudes uncorrected for extinction in the host galaxies.
They also found that, when the magnitudes are corrected for
extinction, SNe with projected separations of less than 10 kpc are 0.3
mag brighter than those farther out. A difficulty in the
interpretation of these results comes about by the use of projected
galactocentric distances, since SNe at large distances can appear
projected near the center of their hosts.  In view of this problem
Ivanov et al. (2000) have recently re-examined the radial distribution
of SNe using {\it deprojected} distances in spiral galaxies, assuming
that SNe Ia come from a disk population.  After normalizing the SN
radial distances by the size of the host galaxies, this study shows no
indication of systematic changes of the SN magnitudes with radial
distance.  Given the abundance gradients observed in spiral galaxies
(\cite{henry99}), this result suggests that the metallicity effects on
SNe Ia are small.

Another tool that can be used to identify the role of age in the SN
properties is provided by the study of SN rates.  According to
Cappellaro et al. (1997) spiral galaxies are 50\% more prolific than
ellipticals in the production of SNe Ia. This result supports the
claim that the SN rate (per unit galaxy luminosity) is enhanced in
spirals, due to a population of SNe with younger progenitors.

In summary, both the theoretical and the observational studies provide
support to the fact that age and metallicity can modify the properties
of SNe Ia. It is unclear yet which of the two is the primary
parameter. Since the theoretical studies have enormous uncertainties
inherent to the physics of flame propagation, it is likely that the
observational approach will provide the clues more quickly.

\begin{figure}
\epsscale{1.0}
\plotone{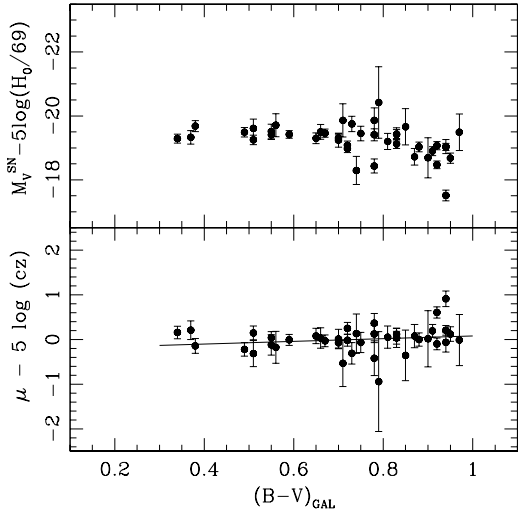}
\figcaption{(top) The absolute $V$ magnitude of the SN
vs. the $(B-V)$ color of the SN host galaxy.  (bottom) Residuals of
the Hubble diagram in the $V$ band vs. the $(B-V)$ color of the SN
host galaxy.\label{fig5}}
\end{figure}

Besides the origin of the SN Ia diversity it is interesting to ask
whether the environmental properties can bias the determination of
distances by application of the M/$\Delta m_{15}(B)$ relation.  Since
the $(B-V)$ color of the SN host galaxies proves to be the best
indicator of the environmental effect on the SN properties, we use this
parameter  in Figure \ref{fig5} to examine this issue. The top panel
shows the absolute $V$ peak magnitude of the SNe Ia versus $(B-V)$ which
confirms the result of Figure \ref{fig2}, i.e., that bluer galaxies on
average produce brighter events. The bottom panel of Figure \ref{fig5}
shows the residuals of the Hubble diagram [$\mu - 5\log(cz) +$
constant] versus $(B-V)$ which reveals that, despite the fact that the
SN peak luminosities vary substantially with galaxy color, the
distances inferred from the M/$\Delta m_{15}(B)$ relation do not depend
on the environmental conditions. The formal fit to the residuals
yields a slope which is only 1$\sigma$ different than zero. At high
redshifts we would expect younger and more metal-poor stellar
populations, and therefore bluer galaxies. If the slight trend shown
in Figure \ref{fig5} was real, it would imply that the distances to
the high-$z$ SNe would be underestimated after application of the
M/$\Delta m_{15}(B)$ relation. A correction for this bias (if any)
would act to increase the distance to the high-$z$ SNe and, thus, to
provide further support to the conclusion that the cosmological
constant is different from zero.

\section{Conclusions}

1) We confirm the result of Hamuy et al. (1996a) that the distribution
of decline rates changes considerably with the morphological type of
the SN host galaxies. We provide further evidence here from a much
larger SN sample and the use of the $(B-V)$ color of the hosts, which
reveals that brighter SNe occur in bluer environments. This suggests
that the progenitor age determines the optical properties of SNe Ia.

2) We find some evidence that bright SNe occur in less luminous galaxies.  In
view that the galaxy luminosities are correlated with galaxy
abundances, this correlation suggests that the diversity of SNe Ia is
due to metallicity variations among the SN progenitors, in the sense
that metal-poor environments produce the most luminous SNe.

3) At the moment it is difficult to separate the effects of ages and
metallicities, because the galaxies of our sample show a correlation
between color and luminosity.

4) A promising route to disentangle age and metallicity effects is to
use spectroscopic data of the SN host galaxies. In our case the five
galaxies with SPP abundances and ages, and the three galaxies with nebular
analysis abundances seem to favor a metallicity effect, although the
low number of objects and the large uncertainties prevent us from
drawing strong conclusions.

5) We use the $(B-V)$ color of the SN host galaxies to quantify
environmental effects in the determination of extragalactic distances
using the M/$\Delta m_{15}(B)$ relation.  Despite the fact that the SN
peak luminosities vary substantially with galaxy color, the distances
inferred from the M/$\Delta m_{15}(B)$ relation do not depend on the
environmental conditions. This results lends credibility to the claim
that the Universe is presently accelerating, due to a non-zero
cosmological constant.

\acknowledgments

We are very grateful to Guy Worthey, D. Arnett, and D. Zaritsky for
useful comments throughout the preparation of this paper.  Support for
SCT was provided by NASA through Hubble Fellowship grant
HF-01125.01-99A awarded by the Space Telescope Science Institute,
which is operated by the Association of Universities for Research in
Astronomy, Inc., for NASA under contract NAS 5-26555.  PAP
acknowledges support from the National Science Foundation through
CAREER grant AST9501634, from the National Aeronautics and Space
Administration through grant NAG~5-2798, and from the Research
Corporation though a Cottrell Scholarship.  This research has made use
of the NASA/IPAC Extragalactic Database (NED) which is operated by the
Jet Propulsion Laboratory, California Institute of Technology, under
contract with the National Aeronautics and Space Administration.

\pagestyle{empty}

\begin{deluxetable} {lclcccccc}
\tablecolumns{9}
\tablenum{1}
\tablewidth{0pt}
\tablecaption{Photometric Observations of SN Ia Host Galaxies}
\tablehead{
\colhead{SN} &
\colhead{$\Delta$m$_{15}$(B)} &
\colhead{Galaxy} &
\colhead{Morph.} &
\colhead{B$_{T}$} &
\colhead{V$_{T}$} &
\colhead{M$_B$-5log(H$_0$/69)} &
\colhead{M$_V$-5log(H$_0$/69)} &
\colhead{B-V} \\
\colhead{} &
\colhead{SN} &
\colhead{} &
\colhead{} &
\colhead{GAL} &
\colhead{GAL} &
\colhead{GAL} &
\colhead{GAL} &
\colhead{GAL} }
\startdata
1937C   &  0.87(10) & IC 4182      & S/Irr      & 11.77(05) & 11.38(12) & -16.65(09) & -17.02(14) & 0.38 \nl
1972E   &  0.87(10) & NGC 5253     & Irr        & 10.87(12) & 10.44(12) & -16.96(16) & -17.34(16) & 0.37 \nl
1980N   &  1.28(04) & NGC 1316     & Sa pec     &  9.42(08) &  8.53(08) & -21.89(10) & -22.76(10) & 0.87 \nl
1981B   &  1.10(07) & NGC 4536     & Sbc        & 11.16(08) & 10.55(08) & -19.86(11) & -20.45(11) & 0.59 \nl
1986G   &  1.73(07) & NGC 5128     & S0+Spec    &  7.84(06) &  6.84(08) & -20.48(09) & -21.37(10) & 0.89 \nl
1989B   &  1.31(07) & NGC 3627     & Sb         &  9.65(13) &  8.92(13) & -20.54(21) & -21.24(21) & 0.70 \nl
1990N   &  1.07(05) & NGC 4639     & Sbc        & 12.24(10) & 11.54(10) & -19.66(14) & -20.34(13) & 0.67 \nl
1990O   &  0.96(10) & PGC 59955    & SBa        & \nodata   & \nodata   & \nodata    &  \nodata   & \nodata \nl
1990T   &  1.15(10) & PGC 63925    & Sa         & 15.61(10) & 14.75(10) & -20.91(15) & -21.66(15) & 0.75 \nl
1990Y   &  1.13(10) & anonymous    & E?         & 16.86(10) & 16.25(10) & -19.37(15) & -19.92(15) & 0.55 \nl
1990af  &  1.56(05) & anonymous    & SB0        & 16.56(10) & 15.45(10) & -20.52(13) & -21.44(13) & 0.92 \nl
1991S   &  1.04(10) & UGC 5691     & Sb         & 15.67(10) & 14.68(10) & -21.56(13) & -22.39(13) & 0.83 \nl
1991T   &  0.94(05) & NGC 4527     & Sbc        & 11.38(08) & 10.52(09) &  \nodata   &  \nodata   & 0.84 \nl
1991U   &  1.06(10) & IC 4232      & Sbc        & 14.73(10) & 13.91(10) & -21.36(17) & -22.09(17) & 0.73 \nl
1991ag  &  0.87(10) & IC 4919      & SBb        & 14.98(10) & 14.34(10) & -19.17(33) & -19.73(33) & 0.56 \nl
1991bg  &  1.93(10) & NGC 4374     & E1         & 10.09(05) &  9.11(05) & -21.33(07) & -22.27(07) & 0.94 \nl
1992A   &  1.47(05) & NGC 1380     & SA0        & 10.87(10) &  9.93(10) & -20.20(14) & -21.12(14) & 0.92 \nl
1992J   &  1.56(10) & anonymous    & E/S0       & 15.67(10) & 14.52(10) & -21.28(14) & -22.22(14) & 0.94 \nl
1992K   &  1.93(10) & ESO 269-57   & SBb        & 12.68(10) & 11.81(10) & -21.19(41) & -21.93(41) & 0.74 \nl
1992P   &  0.87(10) & IC 3690      & SBa        & 15.55(10) & 14.71(10) & -19.90(19) & -20.68(19) & 0.78 \nl
1992ae  &  1.28(10) & anonymous    & Scd/Sd     & 18.08(10) & 17.43(10) & -19.78(12) & -20.27(12) & 0.49 \nl
1992ag  &  1.19(10) & ESO 508-67   & S          & \nodata   & \nodata   & \nodata    &  \nodata   & \nodata \nl
1992al  &  1.11(05) & ESO 234-69   & Sb         & \nodata   & \nodata   & \nodata    &  \nodata   & \nodata \nl
1992aq  &  1.46(10) & anonymous    & Sa?        & 19.09(10) & 17.86(10) & -19.64(11) & -20.55(11) & 0.91 \nl
1992au  &  1.49(10) & anonymous    & E1         & \nodata   & \nodata   & \nodata    &  \nodata   & \nodata \nl
1992bc  &  0.87(05) & ESO 300-9    & Sab        & 15.74(10) & 15.02(10) & -19.09(24) & -19.75(24) & 0.66 \nl
1992bg  &  1.15(10) & anonymous    & Sa         & 16.15(10) & 15.25(10) & -20.65(17) & -21.31(17) & 0.65 \nl
1992bh  &  1.05(10) & anonymous    & Sbc        & 16.57(10) & 16.13(10) & -20.10(14) & -20.44(14) & 0.34 \nl
1992bk  &  1.57(10) & ESO 156-8    & E1         & 14.75(10) & 13.82(10) & -22.61(13) & -23.33(13) & 0.72 \nl
1992bl  &  1.51(10) & ESO 291-11   & SB0/SBa    & 15.52(10) & 14.51(10) & -21.06(14) & -21.95(14) & 0.88 \nl
1992bo  &  1.69(05) & ESO 352-57   & E5/S0      & 14.91(10) & 13.95(10) & -19.78(26) & -20.65(26) & 0.87 \nl
1992bp  &  1.32(10) & anonymous    & S0/Sa      & 17.53(10) & 16.56(10) & -20.70(12) & -21.40(12) & 0.70 \nl
1992br  &  1.69(10) & anonymous    & E0/1       & 18.51(10) & 17.22(10) & -19.96(11) & -20.91(11) & 0.95 \nl
1992bs  &  1.13(10) & anonymous    & SBb        & 16.24(10) & 15.61(10) & -21.14(12) & -21.65(12) & 0.51 \nl
1993B   &  1.04(10) & anonymous    & SBb        & 17.52(10) & 16.80(10) & -20.37(12) & -20.91(12) & 0.55 \nl
1993H   &  1.69(10) & ESO 445-66   & SBb(rs)    & 14.72(10) & 13.82(10) & -20.79(20) & -21.56(20) & 0.78 \nl
1993O   &  1.22(05) & anonymous    & E5/S01?    & 17.62(10) & 16.74(10) & -19.50(13) & -20.22(13) & 0.72 \nl
1993ac  &  1.19(10) & PGC 17787    & E          & \nodata   & \nodata   & \nodata    &  \nodata   & \nodata \nl
1993ae  &  1.43(10) & IC 126       & E          & \nodata   & \nodata   & \nodata    &  \nodata   & \nodata \nl
1993ag  &  1.32(10) & anonymous    & E3/S01     & 16.68(10) & 15.60(10) & -20.65(14) & -21.49(14) & 0.83 \nl
1993ah  &  1.30(10) & ESO 471-27   & SBb        & 15.56(10) & 14.66(10) & -20.13(18) & -20.94(18) & 0.81 \nl
1994D   &  1.32(05) & NGC 4526     & S0         & 10.66(06) &  9.70(06) & -20.29(10) & -21.23(10) & 0.94 \nl
1994M   &  1.44(10) & NGC 4493     & E          & \nodata   & \nodata   & \nodata    &  \nodata   & \nodata \nl
1994Q   &  1.03(10) & PGC 59076    & S0         & \nodata   & \nodata   & \nodata    &  \nodata   & \nodata \nl
1994S   &  1.10(10) & NGC 4495     & Sab        & \nodata   & \nodata   & \nodata    &  \nodata   & \nodata \nl
1994T   &  1.39(10) & PGC 46640    & Sa         & \nodata   & \nodata   & \nodata    &  \nodata   & \nodata \nl
1994ae  &  0.86(05) & NGC 3370     & Sc         & \nodata   & \nodata   & \nodata    &  \nodata   & \nodata \nl
1995D   &  0.99(05) & NGC 2962     & S0         & 12.96(13) & 11.93(13) & -19.88(58) & -20.85(58) & 0.97 \nl
1995E   &  1.06(05) & NGC 2441     & Sb         & 13.00(20) & 12.19(21) & -20.63(42) & -21.41(43) & 0.78 \nl
1995ac  &  0.91(05) & anonymous    & S          & \nodata   & \nodata   & \nodata    &  \nodata   & \nodata \nl
1995ak  &  1.26(10) & IC 1844      & Sbc        & \nodata   & \nodata   & \nodata    &  \nodata   & \nodata \nl
1995al  &  0.83(05) & NGC 3021     & Sbc        & \nodata   & \nodata   & \nodata    &  \nodata   & \nodata \nl
1995bd  &  0.84(05) & UGC 3151     & S          & \nodata   & \nodata   & \nodata    &  \nodata   & \nodata \nl
1996C   &  0.97(10) & MCG +08-25-47& Sa         & \nodata   & \nodata   & \nodata    &  \nodata   & \nodata \nl
1996X   &  1.25(05) & NGC 5061     & E0         & 11.28(13) & 10.36(13) & -21.64(57) & -22.49(57) & 0.85 \nl
1996Z   &  1.22(10) & NGC 2935     & Sb         & 12.10(14) & 11.33(14) & -21.03(52) & -21.74(52) & 0.71 \nl
1996ai  &  0.99(10) & NGC 5005     & SBbc       & 10.61(08) &  9.81(08) & \nodata    &  \nodata   & 0.79 \nl
1996bk  &  1.75(10) & NGC 5308     & S0         & 12.31(15) & 11.39(15) & -20.22(63) & -21.12(63) & 0.90 \nl
1996bl  &  1.17(10) & anonymous    & SBc        & \nodata   & \nodata   & \nodata    &  \nodata   & \nodata \nl
1996bo  &  1.25(05) & NGC 673      & Sc         & 13.20(13) & 12.61(13) & -21.36(30) & -21.88(30) & 0.51 \nl
1996bv  &  0.93(10) & UGC 3432     & S          & \nodata   & \nodata   & \nodata    &  \nodata   & \nodata \nl
1998bu  &  1.01(05) & NGC 3368     & Sab        & 10.11(13) &  9.25(13) & -20.19(16) & -21.03(16) & 0.83 \nl
\enddata
\end{deluxetable}

\pagestyle{empty}

\begin{deluxetable} {llcccccc}
\tablecolumns{8}
\tablenum{2}
\tablewidth{0pt}
\tablecaption{Line Strengths of SN Host Galaxies through r$_e$/8 Aperture}
\tablehead{
\colhead{SN}  &  
\colhead{Galaxy} &
\colhead{H$\beta$}  & 
\colhead{Mg$_2$} &
\colhead{Mgb} &
\colhead{Fe5270} &
\colhead{Fe5335} &
\colhead{$<$Fe$>$} \\
\colhead{   } &
\colhead{   } &
\colhead{(\AA)} &
\colhead{(mag)} &
\colhead{(\AA)} &
\colhead{(\AA)} &
\colhead{(\AA)} &
\colhead{(\AA)} }
\startdata
1991bg   & NGC 4374  & 1.51(04) & 0.243(016) & 4.78(03) & 2.94(04) & 2.69(04) & 2.82(03) \nl
1992A    & NGC 1380  & 2.08(24) & 0.226(023) & 4.15(45) & 2.85(36) & 2.39(42) & 2.62(56) \nl
1994D    & NGC 4526  & 1.62(23) & 0.248(024) & 4.33(47) & 3.00(36) & 2.96(44) & 2.98(57) \nl
1995D    & NGC 2962  & 1.23(26) & 0.263(022) & 3.92(44) & 3.00(36) & 2.62(42) & 2.81(55) \nl
1996X    & NGC 5061  & 2.78(17) & 0.238(015) & 3.50(30) & 2.74(24) & 2.59(30) & 2.67(39) \nl

\enddata
\end{deluxetable}

\begin{deluxetable} {lrrrr}
\tablecolumns{5}
\tablenum{3}
\tablewidth{0pt}
\tablecaption{Ages and Metallicities of the SN Ia Host Galaxies}
\tablehead{
\colhead{SN} &
\colhead{Age} &
\colhead{[Z/H]} &
\colhead{[E/Fe]} &
\colhead{[O/H]} \\
\colhead{} &
\colhead{(Gyr)} &
\colhead{} }
\startdata
1972E   &   \nodata  & \nodata                    & \nodata                    & -1.93(07)\tablenotemark{a} \nl
1991bg  & 12.7(1.9)  & +0.12(03)                  & +0.20(01)                  & \nodata                    \nl
1992A   &  3.3(4.3)  & +0.26(20)                  & +0.23(12)                  & \nodata                    \nl
1994D   &  9.4(8.5)  & +0.16(16)                  & +0.09(11)                  & \nodata                    \nl
1995D   & 24.3(7.2)  & -0.23(12)                  & +0.01(10)                  & \nodata                    \nl
1996X   &  1.4(0.3)  & +0.56(18)                  & +0.20(11)                  & \nodata                    \nl
1996ai  &   \nodata  & \nodata                    & \nodata                    & -0.93(07)\tablenotemark{b} \nl
1998bu  &   \nodata  & \nodata                    & \nodata                    & -1.03(07)\tablenotemark{b} \nl
\tablenotetext{a}{Kobulnicky et al. (1999)}
\tablenotetext{b}{Oey \& Kennicutt (1993)}
\enddata
\end{deluxetable}

\clearpage

\clearpage

\end{document}